\documentclass[10pt,twocolumn,letterpaper]{article}

\usepackage{cvpr} 
\usepackage{times}
\usepackage{epsfig}
\usepackage{graphicx}
\usepackage{amsmath}
\usepackage{makecell}
\usepackage{amssymb}
\usepackage{multirow} 
\usepackage{xcolor}
\usepackage{booktabs}

\begin{document}

\title{LENS-DF: \\ Deepfake Detection and Temporal Localization for Long-Form Noisy Speech}

\author{Xuechen Liu ~ ~ ~ ~ ~ ~ Wanying Ge ~ ~ ~ ~ ~ ~ Xin Wang ~ ~ ~ ~ ~ ~ Junichi Yamagishi \\
National Institute of Informatics, 2-1-2 Hitotsubashi, Tokyo, 101-8430 Japan\\
{\tt\small \{xuecliu, gewanying, wangxin, jyamagis\}@nii.ac.jp}
}

\maketitle
\thispagestyle{empty}

\begin{abstract}
This study introduces \textbf{LENS-DF}, a novel and comprehensive recipe for training and evaluating audio deepfake detection and temporal localization under complicated and realistic audio conditions. The generation part of the recipe outputs audios from the input dataset with several critical characteristics, such as longer duration, noisy conditions, and containing multiple speakers, in a controllable fashion. The corresponding detection and localization protocol uses models. We conduct experiments based on self-supervised learning front-end and simple back-end. The results indicate that models trained using data generated with LENS-DF consistently outperform those trained via conventional recipes, demonstrating the effectiveness and usefulness of LENS-DF for robust audio deepfake detection and localization. We also conduct ablation studies on the variations introduced, investigating their impact on and relevance to realistic challenges in the field\footnote{This study is supported by the New Energy and Industrial Technology Development Organization (NEDO, JPNP22007), and JST AIP Acceleration Research (JPMJCR24U3). This study was partially carried out using the TSUBAME4.0 supercomputer at the Institute of Science Tokyo.}.
\end{abstract}

\section{Introduction}
The exponential growth of audio deepfake technology via various speech synthesis methods, driven by advanced deep learning techniques for realistic generation, has given rise to serious societal concern \cite{brewster2021, Karimi2023, cointelegraph2024}. Deepfake audio clips can spread convincingly fake information, threatening information security. Researchers respond to such challenges by developing various detection methods, leading to adequate deepfake detectors --- usually binary classifiers—that can spot synthetic speech. Advancing beyond conventional classifiers, recent detection models \cite{spoof_cnn_e2e2021, spoof_cnn_lightcnn2020, aasist2022} leveraging deep neural networks (DNNs) have shown promising performance.

In parallel to the advances in spoofing detectors, datasets that are used to train and evaluate them have also advanced. Mainly driven by the challenge series such as ASVspoof \cite{asvspoof2015, asvspoof2019, asvspoof2021_summary} and Audio Deepfake Detection (ADD) \cite{add2022}, these datasets have become complex. Later research has tackled partial spoofing \cite{partialspoof2022}, different compression, including neural-based codecs; and added background noise. Studies have also expanded to languages beyond English \cite{mlaad} and incorporated video alongside audio \cite{cai2023avdeepfake1mlargescalellmdrivenaudiovisual}.

\begin{figure}[t]
    \centering
    \includegraphics[width=\linewidth]{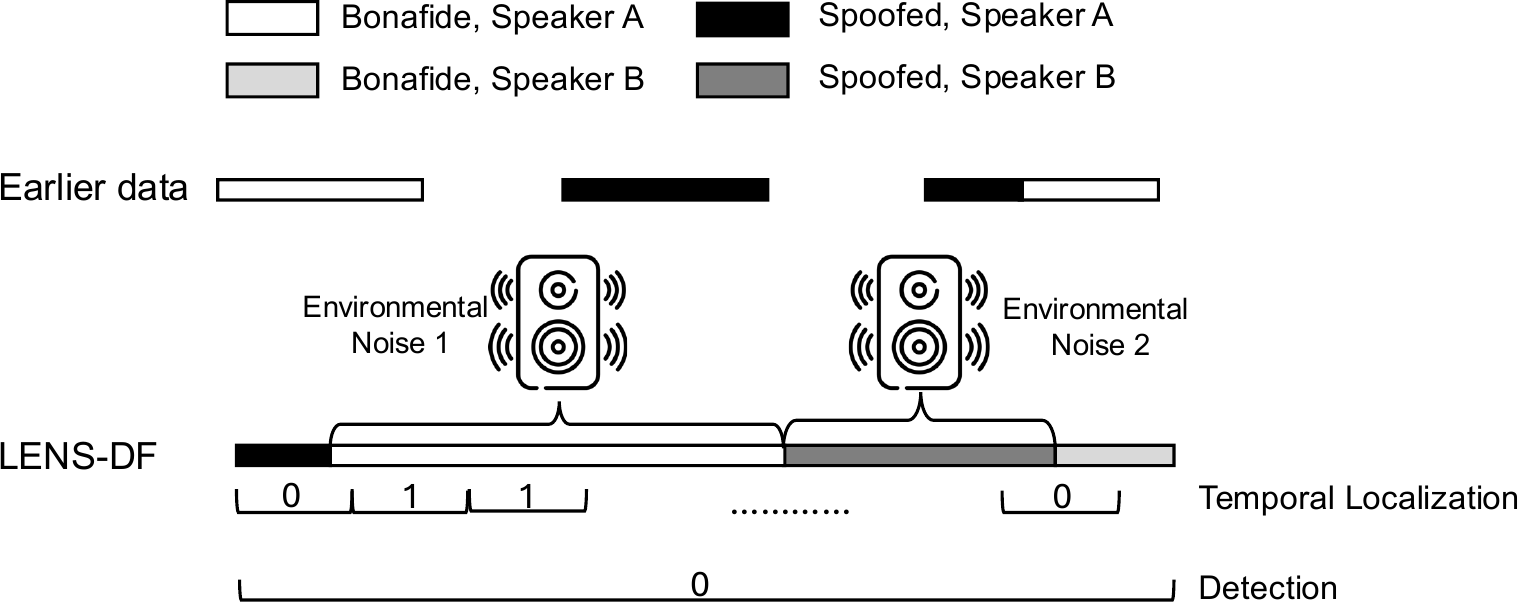}
    \caption{Conceptual differences between earlier datasets and datasets generated for this study. Note that the boxes in the figure represent audio samples. Earlier datasets typically feature short (2–10 seconds), single-speaker audio under clean, stable conditions. Dataset generated with LENS-DF includes longer clips (normally 15-50 seconds), environmental noise, and multiple speakers per sample. Bottom shows the conceptual task definition with how the annotation is conducted (1 denotes bonafide samples, and 0 denotes spoofed ones).}
    \label{fig:novelty_idea}
\end{figure}

Although recent DNN-based deepfake detectors show promising performance, their real-world applicability remains limited because they have been predominantly trained and evaluated under laboratory-controlled audio conditions. State-of-the-art detectors have been trained on typically clean, short segments, ranging from under 2 to approximately 10 seconds \cite{aasist2022, asvspoof2019, rawnet2_spoofing2021}. Real-world issues and factors related to audio deepfakes, such as complex acoustic conditions, extended audio durations, and multi-speaker scenarios, remain underexplored. These issues and factors present additional challenges for audio deepfake detection, highlighting the disparity between the current research and real-world conditions and motivating more representative studies. Addressing these challenges requires new training and evaluation data and a comprehensive methodology covering data generation, training, and evaluation.

To address the current under-exploration in the research field, we introduce LENS-DF (\textbf{L}onger duration, \textbf{E}nhanced multi-speaker, \textbf{N}oisy \textbf{S}peech for audio \textbf{D}eep\textbf{F}ake detection). As shown in Figure \ref{fig:novelty_idea}, it contains several interesting characteristics such as longer duration, noisy conditions, and multi-speaker presence compared with early datasets and recipes. We further test whether state-of-the-art detectors can feasibly generalize in this scenario, measuring performance on both detection and localization. As illustrated at the bottom of the figure, throughout the paper, detection refers to a binary decision at the sample level. Localization, on the other hand, refers to predicting whether a specific region of the audio sample is spoofed.
\begin{itemize}
    \item We introduce LENS-DF, a simple and comprehensive data generation recipe for robust audio deepfake detection that simulates multiple real-world conditions, , including extended audio duration, noise conditions, and multiple speakers  within a single audio sample. We use ASVspoof 2019 logical access (LA)  \cite{asvspoof2019} for this study and generate variants of LENS-DF for training and evaluation\footnote{Data generation code and sample data can be found at \url{https://github.com/nii-yamagishilab/LENS-DF-DataGen}}.
    \item We conduct an investigation into current state-of-the-art neural deepfake detection models based on self-supervised learning (SSL) by fine-tuning them with the training data generated with LENS-DF, examining their adaptability and performance in audio deepfake detection and localization under complex conditions. We also benchmark the performance against various realistic factors and systematically analyze their impact on robustness and reliability.
\end{itemize}

\section{Related Work}
\label{sec:related_work}

\subsection{Audio Deepfake Datasets}
\label{secsec:datasets_review}
Beyond advances in speech synthesis techniques, such as \emph{text-to-speech} and \emph{voice conversion} (VC) for spoofing attacks, recent datasets in audio deepfake research have increasingly emphasized realistic non-speech audio characteristics. ASVspoof 2021 \cite{asvspoof2021_summary} was among the first to address this critical aspect by incorporating audio-compression techniques, including transmission and media codecs, applied to initially clean source audio \cite{asvspoof2019}. Conversely, a different approach was taken with datasets such as In-the-Wild \cite{muller22_inthewild} and DeepFakeEval \cite{deepfakeeval2024} in that audio samples are directly collected from real-world recordings already containing natural artifacts and noise. Another realism-enhancing factor is the inclusion of multiple languages beyond English, as addressed with the Multi-Language Audio Anti-Spoof Dataset \cite{mlaad}. Previous studies (e.g., \cite{xuechen_longform2024}) began exploring spoofing detection under complex conditions and with longer audio durations. However, these initial explorations typically focused on evaluating pre-trained models without SSL front-ends, and the impact of real-world factors has yet to be thoroughly investigated, aspects that this study aims to comprehensively address.

\subsection{Audio Deepfake Detection and Localization}
\label{secsec:models_review}
Audio deepfake detection has been an active research area for decades. Early approaches primarily relied on statistical models, such as \emph{Gaussian mixture models} \cite{gmm_max2018}, combined with various acoustic features \cite{spoof_features2015, cqcc_max}. Advances have seen a shift toward DNN-based methods, including architectures such as RawNet2 \cite{rawnet2, rawnet2_spoofing2021} and AASIST \cite{aasist2022}. Data-augmentation techniques, such as RawBoost \cite{rawboost2022}, have significantly enhanced these models, achieving promising performance under controlled, lab-like evaluation conditions. The evolution of deepfake detectors has increasingly embraced Transformer architectures; for example, \cite{transformer_audio2023_1} integrates a Transformer encoder with a 1-D ResNet to effectively detect partially spoofed audio clips, while RawFormer \cite{rawformer_audi2023_2} combines convolutional and Transformer layers to further refine detection capabilities. Leveraging large-scale pre-trained SSL speech models as front-ends has also proven highly effective. Extracted features from these SSL models can be seamlessly used by downstream classifiers, ranging from simple structures such as global average pooling (GAP) and fully connected (FC) layers \cite{ssl_frontends2022}, to more sophisticated models such as AASIST \cite{tak22_odyssey}.

In parallel to the widely-known detection task, the task of \emph{temporal localization} focuses on locating manipulated segments within an audio utterance, shifting the objective from utterance-level classification toward fine-grained, segment-level detection. Benchmarks such as PartialSpoof \cite{partialspoof2022} and ADD \cite{add2022} have supported research in this area. 
Further, a content-driven partially spoofed audio dataset was developed, in which specific parts of the speech are altered to change the sentential meaning \cite{lav-df, cai2023avdeepfake1mlargescalellmdrivenaudiovisual}. DNN-based methods dominate temporal-localization studies; for instance, an earlier study used hybrid convolutional and recurrent neural networks to effectively capture spectro-temporal patterns and contextual information at the frame level \cite{mt_localization_2021}. Transformer-based methods have been increasingly used \cite{rawformer_audi2023_2, attention_localization_2024}, along with ensemble frameworks combining boundary detection and classification enhanced by SSL front-ends \cite{bounary_localization_2023}. Approaches using temporal convolution paired with contrastive learning have also emerged, further advancing localization performance \cite{contrastive_localization_2024}. However, previous studies primarily evaluated these models using relatively short audio samples under controlled conditions. There has been no exploration of spoof localization for long-duration audio with complex, realistic situations, which is a gap this study aims to address.

\begin{figure*}[t]
    \centering
    \includegraphics[width=0.85\linewidth]{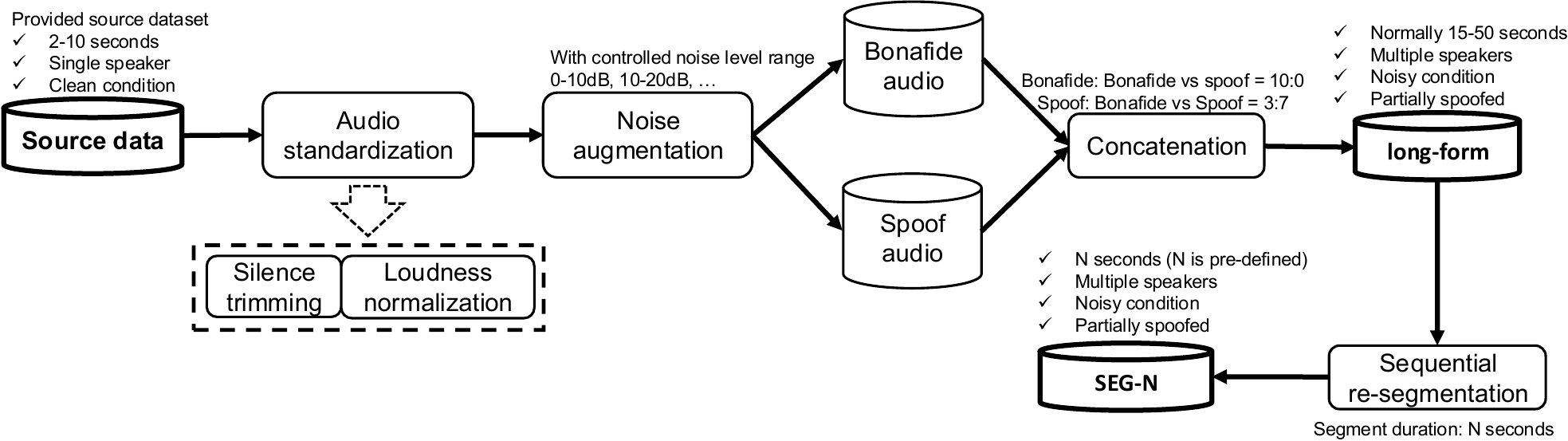}
    \caption{Data-generation pipeline of LENS-DF. Datasets marked as bold-lined cylinders are used to train and evaluate deepfake detectors in this study.}
    \label{fig:generation}
\end{figure*}

\begin{figure*}[t]
    \centering
    \includegraphics[width=0.7\linewidth]{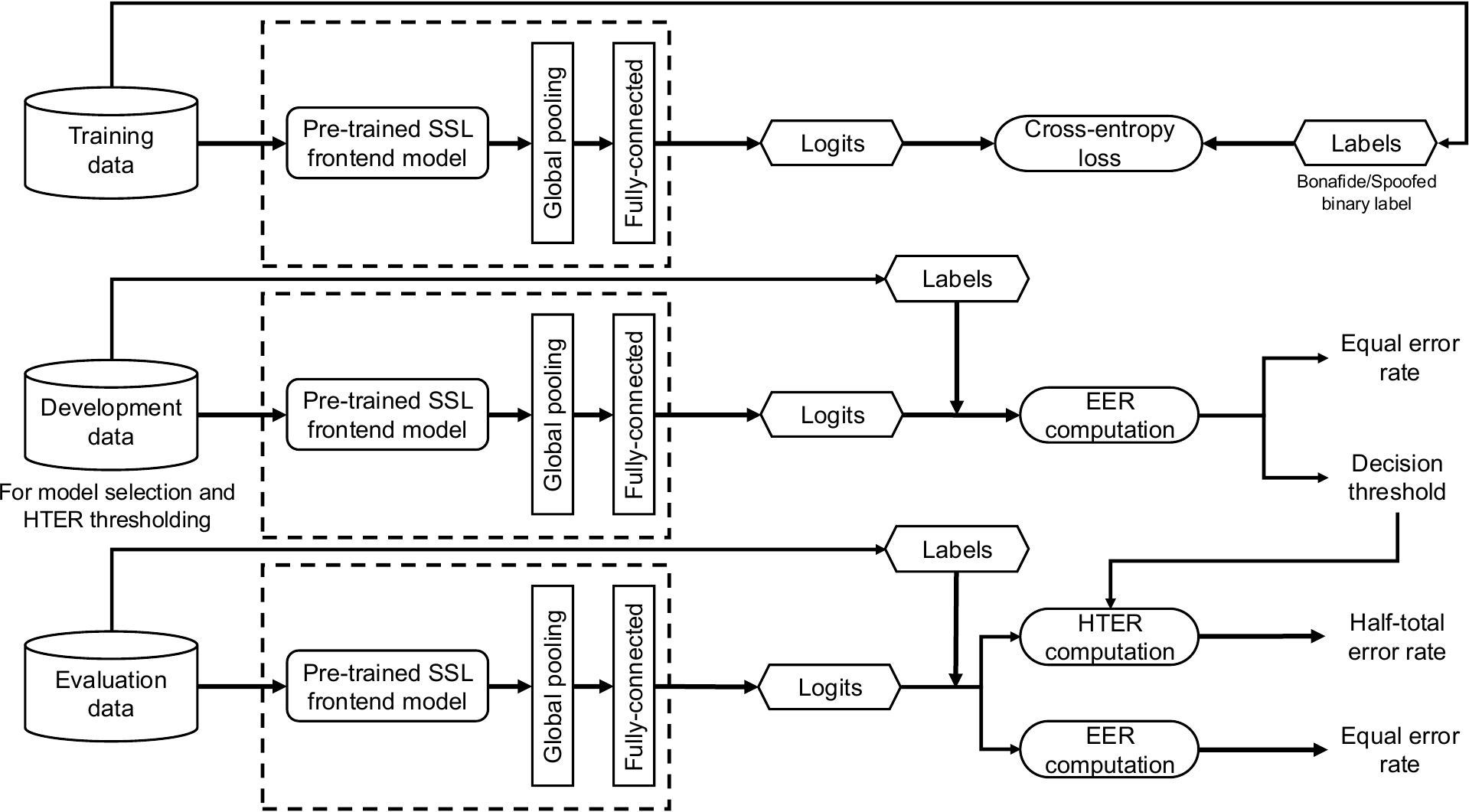}
    \caption{Detection and localization paradigm for LENS-DF. Area within dashed lines shows model architecture used in this study.}
    \label{fig:detection}
\end{figure*}

\section{LENS-DF}
\label{sec:main_dataset}
In this section, we describe the data generation and annotation pipeline and the corresponding detection and localization paradigm of LENS-DF. 

\subsection{Source Data: ASVspoof 2019 LA}
\label{secsec:asvspoof2019la}
As an exemplar source dataset to demonstrate LENS-DF, we use ASVspoof 2019 LA \cite{asvspoof2019}. This dataset is widely used for speech anti-spoofing, including recordings from 107 speakers and spoofed utterances generated using 17 different TTS and VC algorithms (or their hybrids). This diversity of speakers and attack methods provides a rich foundation for creating partial deepfakes. All clips in the dataset are labeled as bonafide or spoofed, and it is essential to note that this dataset does not contain any partially spoofed audio --- each spoofed file is entirely synthetic. The original audio lengths range from 2 to 10 s, and the audio-recording condition is controlled from its source --- the VCTK dataset \cite{vctk}, with a single speaker in each audio sample and clean acoustic condition. This makes this dataset suitable for this study. The original training, development, and evaluation partitions of the dataset\footnote{\url{https://www.asvspoof.org/index2019.html}} [are maintained throughout this paper | were maintained throughout this study?]. The audio throughout this study was mono-channel, with a sampling rate 16KHz.

\subsection{Generation}
\label{secsec:generation}
The data-generation pipeline, depicted in Fig.~\ref{fig:generation}, encompasses several systematic steps designed to produce realistic, standardized, multi-speaker audio under controlled noise conditions\footnote{Note that different from earlier studies, the term ``generation" for this study does not mean speech synthesis via generative models.}. While the pipeline shares some procedural similarities with \cite{xuechen_longform2024}, it significantly differs from the latter in applying more controlled and systematic techniques. The detailed steps are as follows.

\paragraph{Audio standardization}
Similar to \cite{xuechen_longform2024}, initial audio standardization consists of two pre-processing stages: 1) silence trimming, where leading and trailing silence is removed using LibROSA\footnote{\url{https://librosa.org/doc/latest/index.html}}, and 2) loudness normalization, involving randomized audio-volume normalization based on the ITU P.56 standard\footnote{\url{https://github.com/openitu/STL}}. Importantly, these steps do not alter the original labels (bonafide/spoofed) of the data.

\paragraph{Noise augmentation}
Noise augmentation uses background noise from the MUSAN dataset \cite{musan}, categorized into human chatter (\texttt{babble}), music (\texttt{music}), and general environmental sounds (\texttt{noise}). Each audio segment is randomly assigned either no noise or exactly one noise type, with noise intensity uniformly randomized between 0 and 10 dB to reflect realistic scenarios in which background noise is typically lower than the primary speech signal. Unlike \cite{xuechen_longform2024}, the proposed noise augmentation in this study does not include room impulse response \cite{rir} and augmentations using media codecs, such as mp3, to maintain control over noise characteristics.

\paragraph{Audio concatenation}
Following noise augmentation, short audio segments are concatenated to create long-form audio clips, each containing exactly 10 segments. Specifically, fully bonafide long-form audio consists exclusively of bonafide segments. In contrast, partially spoofed audio contains three bonafide and seven spoofed segments in randomized order, based on pilot experiments and \cite{xuechen_longform2024}. To facilitate comparative analysis, the number of generated bonafide and spoofed long-form samples match that in ASVspoof 2019 LA (2,580 bonafide, 22,800 spoofed). Development and evaluation subsets contain 1,000 samples per class, with audio lengths typically 15 to 50 seconds, featuring multiple speakers and varied noise conditions. Labels are assigned on the basis of any spoofed content within the concatenation. We label the data generated from this step as \texttt{long} in the remainder of this paper.

\paragraph{Audio re-segmentation}
We introduce an additional re-segmentation step absent in prior studies \cite{xuechen_longform2024}. Each long-form audio is uniformly segmented into shorter clips of $N$ seconds (default $N=4$), with any residual audio at the end discarded. The motivation for this re-segmentation step is to practically increase the number of training samples under the current training paradigm and enhance the model's capability to detect localized spoofing artifacts, which may occur in specific segmental and temporal regions. Table~\ref{tab:dataset_stats} shows the resulting number of audio samples. Labels follow the same rule as in the last step. The data generated from this step are labeled \texttt{SEG-N} throughout the remainder of this paper.

\subsection{Detection \& Localization}
\label{secsec:detection_localization}
In conjunction with our generation pipeline, we implement the training and evaluation procedures depicted in Figure \ref{fig:detection}. Our reference model uses an SSL front-end to generate frame-level audio representations, subsequently processed by a straightforward back-end classifier composed of global average pooling (GAP) and a fully connected (FC) layer, producing posterior logits for binary classification. During training, model selection is guided by logits computed from the development set on the basis of the equal error rate (EER).

Perhaps it is essential to highlight that our inference strategy differs from conventional video and audio-visual deepfake localization approaches \cite{lav-df, forgerynet}, with which the model produces a boundary-matching confidence map. We conduct localization by creating a ``confidence" per fixed-length segment. Each audio segment, obtained by explicitly re-segmenting long audio recordings with annotations, is independently classified as either bonafide or spoofed. This approach means that longer audio inputs are directly processed with a model trained on shorter fixed-length segments, and each segment receives a separate binary classification without any fusion or aggregation across segments. This aligns with earlier audio-specific deepfake detection and localization practices \cite{mt_localization_2021, partialspoof2022, add2022}, directly yielding segment-level localization decisions.

The optimal threshold identified from the development-set EER is used to calculate the half total equal error rate (HTER) \cite{biometric_evaluation2014}, another widely used metric in biometric evaluations alongside the EER. The EER is computed directly from the logits at evaluation time, while the HTER uses the predefined threshold from the development set. The conditions of the development set matched those of the training set throughout our experiments. Both metrics were used for the two tasks presented in this study.

\begin{table}[t]
\small
\centering
\caption{Number of generated speech samples from ASVspoof 2019 LA, using LENS-DF data generation pipeline. \texttt{SEG-4} samples are \emph{4-s} and from \texttt{long}.}
\begin{tabular}{c|cc|cc}
\hline
 & \multicolumn{2}{|c|}{\texttt{long}} & \multicolumn{2}{c}{\texttt{SEG-4}} \\ \hline
\textbf{Partition} & \textbf{Bonafide} & \textbf{Spoofed} & \textbf{Bonafide} & \textbf{Spoofed} \\ \hline
Train & 2,580 & 22,800 & 17,857 & 129,805 \\ 
Dev & 1,000 & 1,000 & 5,132 & 5,640 \\
Eval & 1,000 & 1,000 & 4,984 & 5,663 \\ \hline
\end{tabular}
\label{tab:dataset_stats}
\end{table}

\section{Experimental Design}
\label{sec:experimental_setup}

\subsection{Training}
\label{secsec:training_setup}
\paragraph{SSL front-end}
On the basis of preliminary experiments, we selected two pre-trained SSL models from the \emph{massive multilingual speech} (MMS) framework \cite{pratap2024_mms}. Both selected models leverage the Wav2Vec 2.0 architecture \cite{baevski2020_wav2vec} and have been pre-trained extensively on speech data covering over 1,400 languages. Specifically, we used two MMS variants: one with around 1 billion parameters (MMS-1B\footnote{\url{https://huggingface.co/facebook/mms-1b}}) and the other with around 300 million parameters (MMS-300M\footnote{\url{https://huggingface.co/facebook/mms-300M}}), with their respective pre-trained checkpoints publicly accessible. The basic information of these two models is listed in Table \ref{tab:ssl_model_info}.

\paragraph{Training details}
As shown in Figure \ref{fig:detection}, we fine-tune the SSL front-end together with the back-end using a cross-entropy loss. For comparison, we experiment with three different training sets for fine-tuning the SSL models: 19LA$_\text{train}$, the original ASVspoof 2019 LA training dataset; \texttt{long}$_\text{train}$, the training partition of our generated data (see Table \ref{tab:dataset_stats}); and \texttt{SEG-4}$_\text{train}$, the re-segmented version of \texttt{long}$_\text{train}$. To improve training efficiency, we use dynamic batching\footnote{\url{https://speechbrain.readthedocs.io/en/latest/tutorials/advanced/dynamic-batching.html}}, enabling input audio durations to vary while remaining within a similar numerical range. This approach also minimizes the need for excessive zero padding. We also constrain the total audio duration per mini-batch to 100 s, ensuring workable memory usage. The temporal pooling via the GAP layer is done at the whole audio length without any further fixed-length chunking. All models are optimized using the Adam optimizer \cite{adam}. The learning rate is linearly increased to $1\times10^{-7}$ over the first 80k steps then linearly decayed to zero by 800k steps, at which point training terminates. We stop training if either 100 epochs or 800k steps are reached, whichever criterion is met first. Each model is trained and evaluated on a single NVIDIA TESLA A100 GPU card.

\begin{table}[t]
\centering
\scriptsize
\caption{Details of two SSL models used for this study. Information is obtained from \cite{pratap2024_mms}.}
\begin{tabular}{c|cccc} 
\hline
    Model & Backbone & Training data & Model size \#params. \\ \hline
    MMS-1B & Wav2Vec 2.0 \cite{baevski2020_wav2vec} & \makecell[c]{55K hours\\ 1K+ languages} & $\sim$965 million  \\ \hline
    MMS-300M & Wav2Vec 2.0 \cite{baevski2020_wav2vec} & \makecell[c]{55K hours\\ 1K+ languages} & $\sim$317 million \\ \hline
\end{tabular}
\label{tab:ssl_model_info}
\end{table}

\subsection{Evaluation}
\label{secsec:evaluation}
\paragraph{Data}
We evaluate both detection and segment-level temporal localization performance on the LENS-DF evaluation set, which comprises our generated long-form audio, as detailed in Section \ref{sec:main_dataset}. We cover the source evaluation data, denoted as ``19LA" in the next section, and the two generated variants via LENS-DF: (i) \texttt{long}, which refers to the generated long-form audio; and (ii) \texttt{SEG-4}, which refers to the re-segmented samples (see Section \ref{secsec:generation}). The length is set to 4 s. We mark the training and evaluation sets using subscripts ``$_\text{train}$" and ``$_\text{eval}$", respectively. We assess performance under both matched-condition and cross-condition settings, where matched conditions use the same variant (including source evaluation data) for training and evaluation. At the same time, cross-conditions involve different variants or sources between the two.

\paragraph{Metrics}
As introduced in Section \ref{secsec:detection_localization}, we use the EER and HTER as the primary evaluation metrics. Following prior research, we treat localization as detection applied to very short segments, computing both metrics in the same manner for segment-based decisions. The HTER threshold is determined using the development set corresponding to the evaluation set to ensure a fair comparison of model performance. For instance, if the evaluation condition is \texttt{long}, the development set will also be from \texttt{long}.

\begin{table*}[th]
\small
\centering
\caption{Statistical results on LENS-DF evaluation set. Note that models in this table do not have RawBoost \cite{rawboost2022} applied. Numbers in \textbf{bold} and \underline{underlined} indicate best and second-best results under each evaluation condition, respectively.}
\begin{tabular}{l|c|cc|cc|cc}
\hline
& & \multicolumn{2}{c|}{\textbf{Detection}, 19LA$_\text{eval}$} & \multicolumn{2}{c|}{\textbf{Detection}, \texttt{long}$_\text{eval}$} & \multicolumn{2}{c}{\textbf{Localization}, \texttt{SEG-4}$_\text{eval}$} \\ \hline
\textbf{SSL Model} & \textbf{Training Data} & \textbf{EER (\%)} & \textbf{HTER (\%)} & \textbf{EER (\%)} & \textbf{HTER (\%)} & \textbf{EER (\%)} & \textbf{HTER (\%)} \\ \hline
MMS-1B & 19LA$_\text{train}$ & \underline{0.39} & \underline{1.97} & 6.40 & 9.30 & 22.10 & 23.04 \\
MMS-300M & 19LA$_\text{train}$ & \textbf{0.15} & \textbf{0.52} & 2.90 & 4.05 & 21.12 & 21.09 \\ \hline
MMS-1B & \texttt{long}$_\text{train}$ & 5.15 & 4.97 & \underline{0.90} & \underline{1.65} & 19.20 & 17.86 \\
MMS-300M & \texttt{long}$_\text{train}$ & 7.45 & 5.32 & 1.30 & \textbf{1.40} & 17.81 & 17.26 \\ \hline
MMS-1B & \texttt{SEG-4}$_\text{train}$ & 4.62 & 7.98 & \textbf{0.60} & 4.20 & \underline{15.14} & \underline{14.34} \\
MMS-300M & \texttt{SEG-4}$_\text{train}$ & 4.92 & 4.66 & 1.00 & 8.40 & \textbf{14.62} & \textbf{14.08} \\ \hline
\end{tabular}
\label{tab:lensdf_results}
\end{table*}

\begin{table*}[th]
\small
\centering
\caption{Statistical results on LENS-DF evaluation set, with RawBoost \cite{rawboost2022} applied as online data-augmentation method during training. The last row is model released by \cite{tak22_odyssey}, which has been pre-trained on 19LA with AASIST back-end and RawBoost applied. \textbf{bold} and \underline{underlined} indicate best and second-best results under each evaluation condition, respectively.}
\begin{tabular}{c|c|c|cc|cc|cc}
\hline
& & & \multicolumn{2}{c|}{\textbf{Detection}, 19LA$_\text{eval}$} & \multicolumn{2}{c|}{\textbf{Detection}, \texttt{long}$_\text{eval}$} & \multicolumn{2}{c}{\textbf{Localization}, \texttt{SEG-4}$_\text{eval}$} \\ \hline
\textbf{SSL Model} & \textbf{Training Data} & \textbf{RawBoost} & \textbf{EER (\%)} & \textbf{HTER (\%)} & \textbf{EER (\%)} & \textbf{HTER (\%)} & \textbf{EER (\%)} & \textbf{HTER (\%)} \\ \hline
\multirow{4}{*}{\centering MMS-300M} & 19LA$_\text{train}$ & Yes & \underline{0.46} & 7.80 & 9.60 & 14.80 & 26.32 & 27.75 \\ 
& \texttt{long}$_\text{train}$ & Yes & 12.06 & 7.99 & \underline{0.90} & \textbf{1.90} & \underline{18.89} & \underline{18.36} \\
& \texttt{SEG-4}$_\text{train}$ & Yes & 8.31 & \underline{6.92} & \textbf{0.60} & \underline{3.80} & \textbf{13.68} & \textbf{13.52} \\ \cline{2-9} 
& \texttt{SEG-4}$_\text{train}$ (Clean) & Yes & 6.85 & 7.68 & 5.90 & 12.30 & 26.66 & 26.65 \\
& \texttt{SEG-4}$_\text{train}$ (Clean) & No & 3.26 & 8.36 & 8.90 & 9.75 & 29.94 & 31.41 \\ \hline \hline
XLS-R-300M & 19LA$_\text{train}$ & Yes & \textbf{0.19} & \textbf{0.94} & 15.70 & 17.60 & 30.41 & 27.30 \\ \hline
\end{tabular}
\label{tab:ablation_da_results}
\end{table*}

\section{Results and Discussion}
\label{sec:results}
The detection and localization results on the LENS-DF evaluation set are presented in Table \ref{tab:lensdf_results}. We have several key findings, which are detailed below.

\textbf{Conventional datasets with short, clean speech are clearly not suitable for training models for detecting long and complex spoofed audios}. Models trained on conventional short, clean speech (ASVspoof 2019 LA) do not perform well when evaluated on \texttt{long}$_\text{eval}$. While MMS-300M achieved the best detection performance on the matched 19LA$_\text{eval}$, its performance significantly degraded under more complex conditions such as \texttt{long}$_\text{eval}$ (from 0.15\ to 2.9\% EER, from 0.52\% to 4.05\% HTER). Conversely, training on \texttt{long}$_\text{train}$ and its re-segmented variant \texttt{SEG-4}$_\text{train}$ improves the performance for both types of models, with MMS-1B being the second-best considering both metrics (0.9\% EER, 1.65\% HTER) and MMS-300M reaching the best EER with the latter training data (0.6\%). These findings highlight the importance of the incorporated complexity required to achieve robust spoofing detection. The complexity in \texttt{long} includes several variations introduced in a compounding manner. Thus, isolating specific factors contributing to improvement is important.

\textbf{Model trained on conventional datasets with short, clean speech needs further effort for localizing long-form deepfake speech in noisy, complex environments}. We now move on to the \texttt{SEG-4}$_\text{eval}$ results. Localization using models trained on 19LA$_\text{train}$ resulted in unsatisfying performance, indicating that models trained solely on short and clean audio without multi-speaker presence struggle with temporal localization. Expectedly, models trained explicitly on the long-form or re-segmented datasets demonstrated improved temporal localization performance, with the best result (MMS-300M trained on \texttt{SEG-4}$_\text{train}$) achieving 14.62\% EER and 14.08\% HTER. 

\textbf{Detection models trained on complex data might not perform localization effectively}. 
Although LENS-DF variants (\texttt{long} and \texttt{SEG-4}) notably improve detection of spoofed speech under complex conditions, localization performance remains inadequate for practical use, with the lowest EER and HTER observed at 14.62\% and 14.08\%, respectively. Although training with \texttt{SEG-4}$_\text{train}$ provided more shorter samples and greater exposure to spoofed content and results in notable improvements compared with \texttt{long}$_\text{train}$, its performance may not be considered feasibly satisfying. These findings highlight the need for distinct training strategies and datasets tailored explicitly. It also indicates the significant impact of the realistic and complex factors introduced in the data-generation pipeline. We discuss these factors via multiple self-contained and detailed ablation studies covered in Section \ref{sec:ablation}.

\section{Ablation Study}
\label{sec:ablation}
In this section, we investigate several realistic factors created in the pipeline illustrated in Figure \ref{secsec:generation}. Specifically, we examine the impact of noise, duration, and presence of multiple speakers within single audio sample. 

\begin{figure}[th]
    \centering
    \includegraphics[width=\linewidth]{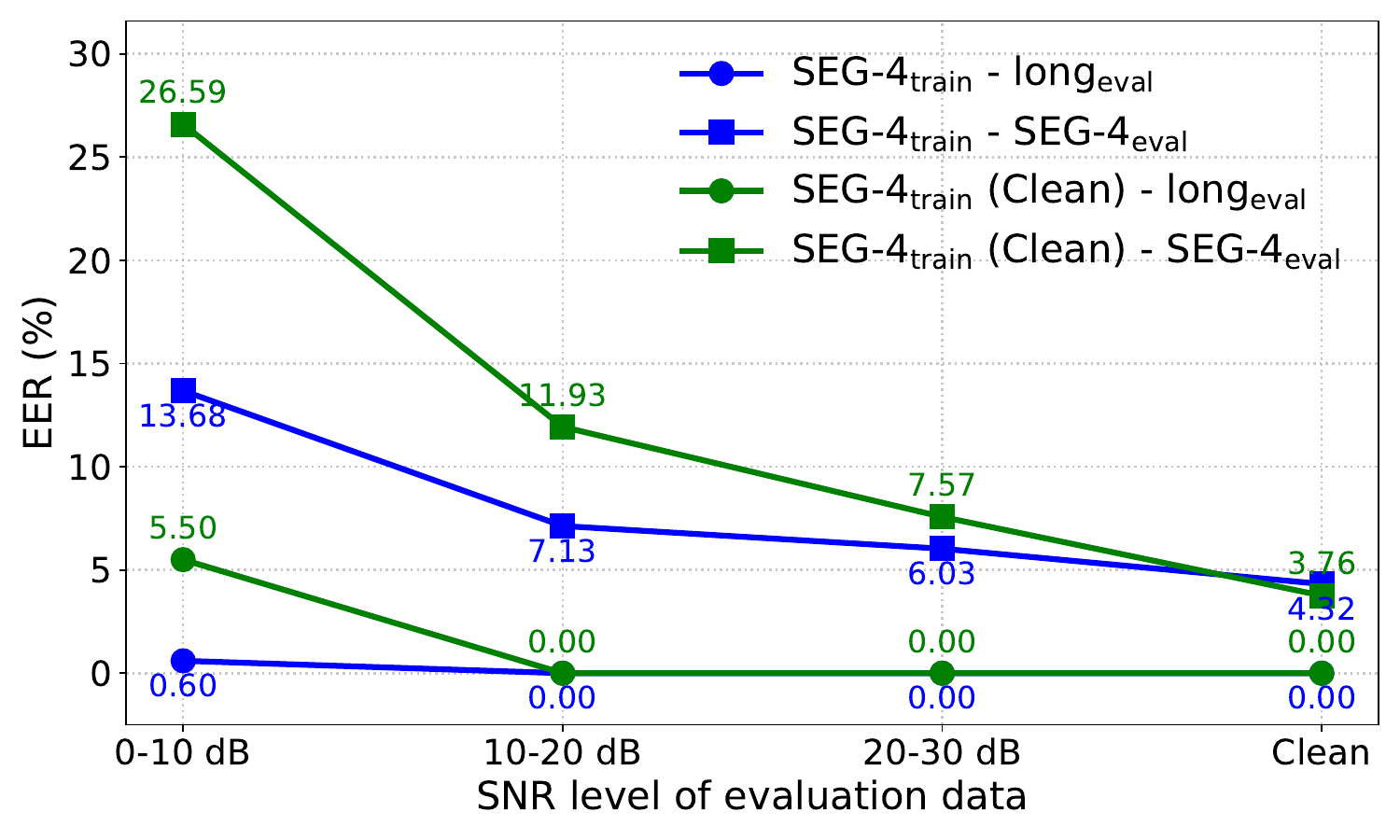}
    \caption{Detection and localization performance against different noise level ranges of LENS-DF evaluation. The difference in colors and shapes corresponds to different training and evaluation conditions, respectively.}
    \label{fig:noise_ablation}
\end{figure}

\subsection{The effect of noise augmentation}
\label{secsec:noise_ablation}
We first study the impact of our controlled noise augmentation during both training and evaluation. Two specific aspects are addressed separately: 1) the comparative effectiveness of our noise augmentation strategy introduced in Section \ref{secsec:generation} versus earlier established methods; and 2) the impact of varying noise levels during evaluation, measured using the signal-to-noise ratio (SNR).

For the first aspect, we employ RawBoost \cite{rawboost2022}, a validated data augmentation method known for controlled additions such as room impulse response \cite{rir} and digital compression using signal processing and filtering techniques. We integrate this augmentation during the data loading phase of training, using a publicly available implementation\footnote{\url{https://github.com/TakHemlata/RawBoost-antispoofing}}.
To systematically evaluate the efficacy of RawBoost in our experimental setup, we first create a variant of \texttt{SEG-4} by removing the noise augmentation step illustrated in Figure \ref{fig:generation}. The variant audio samples retain longer duration and multi-speaker presence in such cases. . We label this variant as \texttt{SEG-4} (Clean) and use it to serve training and evaluation phases. We then examine the performance impact of applying RawBoost on \texttt{SEG-4}$_\text{train}$ (Clean), allowing for a direct and fair comparison against the noise-augmentation strategy we introduced in our study.
For additional comparative reference, we evaluate a baseline model comprising an XLS-R \cite{xlsr} front-end with approximately 300 million parameters and an AASIST back-end \cite{aasist2022}. This baseline, labeled as XLS-R-300M, was trained using randomly chunked or padded audio segments of 4 seconds in duration. The pre-training is carried out on the 19LA$_\text{train}$ dataset with RawBoost applied, achieving a reported 0.82\% EER on the ASVspoof 2021 LA evaluation set \cite{tak22_odyssey}\footnote{Checkpoint available at \url{https://github.com/TakHemlata/SSL_Anti-spoofing}}.

Results are presented in Table \ref{tab:ablation_da_results}. In comparison to findings from Table \ref{tab:lensdf_results}, for more complex training scenarios (such as those utilizing \texttt{long} or \texttt{SEG-4} data), RawBoost provides moderate improvements, especially in the \texttt{long}$_\text{eval}$ scenario. This gain comes at a considerable performance penalty for the 19LA detection task (where EER deteriorates from 7.45\% to 12.06\%). Localization improvements under these conditions are modest, with EER improving slightly from 14.62\% to 13.68\%.
Removing our noise augmentation from \texttt{SEG-4}$_\text{train}$ notably degrades performance on the two complex evaluation sets, with moderate improvement at 19LA$_\text{eval}$ in terms of EER. Applying RawBoost recovers such degradation for \texttt{long}$_\text{eval}$ detection but does not for the other two sets. These results collectively suggest that RawBoost demonstrates limited effectiveness compared with LENS-DF due to its highly controlled setup.

Regarding the second aspect, we examine the impact of varying evaluation noise levels, considering practical conditions with non-negative SNRs. Given that our training data is augmented within an SNR range of 0 to 10dB, we evaluate model performance across evaluation sets with noise levels ranging from 0 to 30 dB at selected intervals. For this analysis, we choose the MMS-300M model trained on \texttt{SEG-4}$_\text{train}$ with RawBoost, as it yielded the best EER and HTER on \texttt{SEG-4}$_\text{eval}$.

The resulting detection and localization performances across these varying SNR conditions on the LENS-DF dataset are illustrated in Figure~\ref{fig:noise_ablation}. Our analysis reveals a clear trend: reducing noise levels enhances detection and localization accuracy. Under noise-free conditions, detection accuracy approaches near-perfect performance (0\% EER), and localization accuracy also significantly improves, achieving around 5\% EER. Conversely, performance deteriorates sharply at higher noise levels, approaching speech signal amplitude. These findings highlight the importance of maintaining a complex yet controlled acoustic environment for reliable detection and temporal localization. Future work may involve exploring more adaptive noise augmentation methods that better align with complex real-world scenarios.

\begin{figure}[t]
    \centering
    \includegraphics[width=\linewidth]{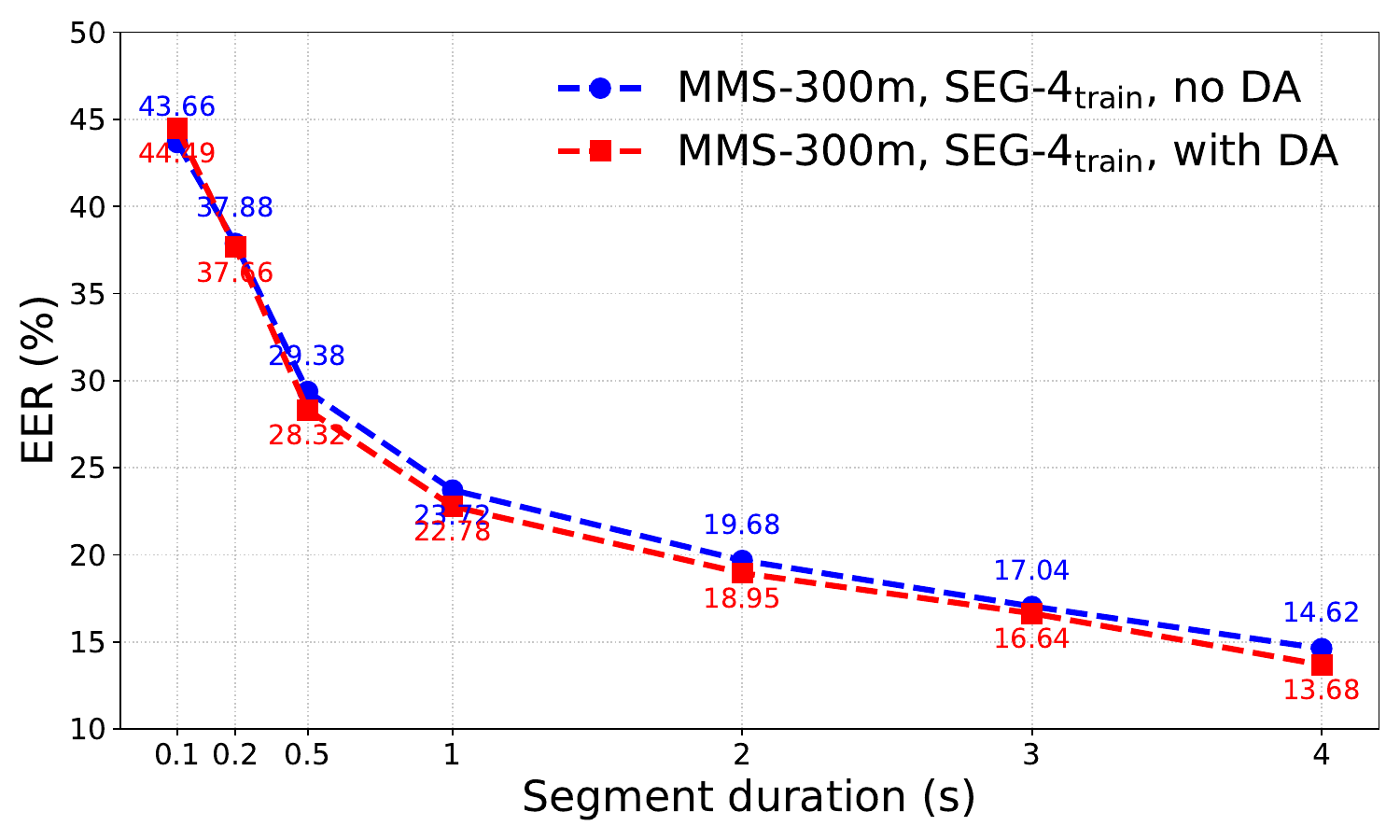}
    \caption{Localization performance against different durations of LENS-DF evaluation set in EER(\%). ``DA" corresponds to the application of RawBoost addressed in Section \ref{secsec:noise_ablation}.}
    \label{fig:duration_ablation}
\end{figure}

\subsection{The effect of duration on localization}
\label{secsec:duration_ablation}
In Section \ref{sec:results}, we explored two different tasks under two distinct segment durations (long-form audio and 4-second segments), observing significant performance differences. We conduct experiments to systematically study this effect by varying the duration ($N$, measured in seconds) of segments re-extracted from the \texttt{long} portion of the LENS-DF evaluation set. Specifically, we vary $N$ from 0.01 (typical hop size of audio framing) up to 4, selecting several intermediate durations for comparison. We use the MMS-300M trained on \texttt{SEG-4} audio, with and without RawBoost applied, as they achieved the best performance on \texttt{SEG-4}$\_\text{eval}$ among all trained models.

The resulting performance is shown in Figure~\ref{fig:duration_ablation}. Extremely short segments yield insufficient localization performance. However, performance consistently improves as segment duration increases. This indicates that longer segments facilitate better temporal localization under current evaluation conditions and setup. Future work may focus on developing architectures and training strategies specifically optimized for ultra-short audio segments.

\begin{table}[t]
\centering
\caption{Detection and localization performance measured using EER (\%) under single- and multi-speaker conditions. \texttt{multi.} in this table is same as \texttt{long}$_\text{train}$ in earlier sections.}
\resizebox{\columnwidth}{!}{
\begin{tabular}{c|cc|cc}
\hline
\multirow{3}{*}{\textbf{Train / Eval cond.}}& \multicolumn{2}{c|}{\texttt{single}} & \multicolumn{2}{c}{\texttt{multi.}} \\ \hline
& \textbf{Detection}, &\textbf{Localization}, & \textbf{Detection}, &\textbf{Localization}, \\
& \texttt{long}$_\text{eval}$ & \texttt{SEG-4}$_\text{eval}$ &\texttt{long}$_\text{eval}$ & \texttt{SEG-4}$_\text{eval}$\\ \hline
\texttt{single}. & 7.90 & 16.10 & \textbf{1.30} & 19.56 \\ \hline
\texttt{multi}. & 11.10 & 17.37 & \textbf{0.60} & \textbf{13.68} \\ \hline
\end{tabular}
\label{tab:single_multi_speaker}
}
\end{table}

\subsection{The effect of speaker presence}
\label{secsec:speaker_ablation}

\begin{table*}[th]
\small
\centering
\caption{Statistical performance of models trained on LENS-DF and LAV-DF, measured by precision and recall metrics. Time resolution was 0.04 seconds.}
\begin{tabular}{c|cccc|cccc}
\hline
\textbf{Data} & \textbf{AP@0.25} & \textbf{AP@0.5} & \textbf{AP@0.75} & \textbf{AP@0.95} & \textbf{AR@100} & \textbf{AR@50} & \textbf{AR@20} & \textbf{AR@10} \\
\hline
LAV-DF & 99.32 & 29.28 & 0.16 & 0.00 & 27.77 & 26.27 & 24.63 & 23.73 \\ \hline
LENS-DF (ours) & 100.00 & 33.50 & 0.00 & 0.00 & 38.56 & 37.52 & 36.61 & 35.70 \\ \hline
\end{tabular}
\label{tab:lav_df_results}
\end{table*}

We then evaluate whether multiple speakers in long-form audio segments can affect detection and localization performance. Similar to Section~\ref{secsec:noise_ablation}, we modify the original data generation pipeline by limiting post-augmentation concatenation to segments from a single speaker (as illustrated in Figure~\ref{fig:novelty_idea}, where only speaker A or B is used). This variant preserves the long-form structure and diverse noise conditions but includes only one speaker per audio. To enable this single-speaker setup, we leverage the speaker correspondence between bonafide and spoofed samples in ASVspoof 2019 LA. We refer to the original multi-speaker variant as \texttt{multi.} and the new single-speaker variant as \texttt{single}. For comparison, we additionally train a model on the \texttt{single} speaker data using the same MMS-300M setup as in Section~\ref{secsec:noise_ablation}.

The results are presented in Table~\ref{tab:single_multi_speaker}. While both models perform well for detection under multiple-speaker conditions (1.3\% and 0.6\% EER), performance notably drops for single-speaker cases (7.9\% and 11.1\% EER). Both models struggle with the \texttt{SEG-4}$_\text{eval}$ regardless of speaker configuration, reflected by the lowest EER being larger than 13\%. These findings suggest that when only one speaker is present, the model may over-rely on speaker identity rather than spoofing artifacts, which leads to shortcut learning and degraded performance, especially with unseen or multi-speaker inputs. Hence, speaker information may be an unreliable cue in this context, and future work may focus on alternative features and strategies to improve robustness.

\subsection{Comparative results on LAV-DF}
\label{secsec:lav_df}
Finally, we compare our dataset with existing work on audio-visual deepfake localization, represented by LAV-DF \cite{lav-df}. We use the data protocol and adopt the evaluation metric from \cite{lav-df, forgerynet} for a sensibly fair comparison. 

It is worth noting that our model hypothesis differs fundamentally from the previous ones incorporating visual information. The audio-visual deepfake detectors trained in those works predict variable-length segments in time, whereas the audio deepfake models in this study work on fixed-length windows. This mismatch affects not just what the model outputs but also how one matches predictions to ground truth. Intersection-over-union (IoU) is typically computed over continuous frame segments and assigns each ground-truth event to its best-matching prediction. The fixed-window approach in audio tasks generates overlapping predictions requiring IoU computation over discrete chunks. Therefore, the matching logic will also change: instead of one prediction per event, we aggregate point-based model predictions with a specific time resolution and pre-defined IoU threshold. Prediction scores for these fixed segments are binarized for classification, and IoU is computed directly from these binary predictions and ground-truth labels per segment. Subsequently, average precision is calculated across varying IoU thresholds, which resembles earlier practices in audio deepfake localization \cite{range_eer_local, old_range_eer_local}. The time resolution is set to 0.04 s, which is the same as that used in a previous study \cite{lav-df}\footnote{In \cite{lav-df}, the audio features used is log spectrogram with 0.01-second resolution, and the scores are downsampled 4 times.}. There is no trimming or padding during our evaluation.
We evaluate using MMS-300M trained on \texttt{SEG-4}$_\text{train}$ without RawBoost. 

As shown in Table \ref{tab:lav_df_results}, model performance on LENS-DF is comparable to that on LAV-DF in AP, despite LAV-DF being more mismatched to our model.
Additionally, our model achieves better AR on LENS-DF, likely due to a more balanced representation of bonafide and spoofed data than on LAV-DF. This observation aligns with previous findings \cite{xuechen_longform2024}, indicating that overly abundant bonafide data may obscure spoofed regions, negatively impacting spoof detection. These results underscore the effectiveness of LENS-DF and the necessity of adapting evaluation metrics specifically tailored to the characteristics of audio-based detection models. Future research could further explore tailored training methods and evaluation frameworks to address better temporal localization challenges in audio-visual tasks represented by LAV-DF and others \cite{cai2023avdeepfake1mlargescalellmdrivenaudiovisual}.

\section{Conclusion and Future Work}
\label{sec:conclusion}
In this study, we have introduced LENS-DF, a comprehensive approach addressing realistic complexities in audio deepfake detection and localization, such as extended audio duration, varying noise conditions, and multi-speaker scenarios within single audio samples. We have extended previous work on audio deepfake data generation in a controllable manner to facilitate both training and evaluation. 
Moreover, we have confirmed through systematic experiments and ablation studies that current state-of-the-art SSL models trained and, while effective under laboratory conditions, demonstrate performance degradation in detection and temporal localization under complex conditions. This degradation can be mitigated by explicitly incorporating additional complexities into the training. Meanwhile, robust localization may require more effort in complex, real-world scenarios, even after systematically exploring multiple contributing factors. Future work will focus on advanced model architectures and training methodologies for temporal localization. Adapting to multi-speaker scenarios, language variations, and noise can further bridge the gap between experimental settings and practical applications.

{\small
\bibliographystyle{ieee}
\bibliography{egbib}
}

\end{document}